\documentclass[journal]{IEEEtran}

\usepackage{amsmath,amssymb}
\usepackage{graphicx}
\usepackage{booktabs}
\usepackage{cite}
\newcommand{\revone}[1]{#1}
\newcommand{\revtwo}[1]{#1}

\graphicspath{{figures/}}

\begin{document}

\title{Fast Whittle Maximum-Likelihood Parametric Spectrum Estimation for Weather-Radar Doppler Moments}

\author{Tworit Dash,~\IEEEmembership{Member,~IEEE,} and Alexander Yarovoy,~\IEEEmembership{Fellow,~IEEE}%
\thanks{This work was partly financially supported by the Dutch Research Council (NWO) under contract WI-NC 175.2021.038.}
\thanks{T. Dash and A. Yarovoy are with the Microwave Sensing, Signals and Systems group, Delft University of Technology, Delft, The Netherlands.}
}

\maketitle

\begin{abstract}
The computational cost of Whittle maximum-likelihood parametric spectrum estimation (PSE) for short-dwell weather-radar Doppler moment retrieval is addressed. PSE can reduce finite-sample bias, especially for short coherent processing intervals and broad spectra, but repeated likelihood evaluations make direct use expensive. A fast implementation of the finite-sample Gaussian Doppler model is introduced by writing the likelihood in terms of closed-form autocorrelation lags and synthesizing the model spectrum and analytic derivatives by fast Fourier transforms (FFTs) on the same $N$-point grid as the periodogram. The estimator is extended to jointly estimate total power, normalized mean Doppler frequency, and normalized spectral width. Monte Carlo simulations with 1000 noise realizations show that the fast PSE keeps the accuracy advantage of the original PSE, while objective-level timing confirms the reduction from dense lag-to-frequency synthesis to FFT-based evaluation.
\end{abstract}


\section{Introduction}
Weather radars commonly estimate the three Doppler moments, namely received power, mean Doppler velocity, and spectral width, from a short time series at each range gate \cite{Doviak1993}. \revone{Classical estimators are still widely used because they are simple and fast. The discrete Fourier transform (DFT) moment method estimates the moments by summing the observed Doppler spectrum and is easy to interpret, but its accuracy depends on spectral resolution, windowing, leakage, and noise correction \cite{Sirmans1975,Dias2000NonparametricWidth}. Pulse-pair (PP), multi-lag, generalized multi-lag, and hybrid spectral-width estimators use one or more sample autocorrelation lags and are computationally efficient for mean velocity and narrow spectra \cite{Zrnic1977,Mahapatra1983,Baxa,Meymaris2009PerformanceORDA,Warde2023GeneralizedObservations}. These multi-lag and hybrid methods can be interpreted as heuristic combinations of lag-domain moment estimates, or as low-dimensional covariance-domain approximations to the Doppler spectrum, but they do not evaluate the finite-sample parametric likelihood and their width estimates can be biased when the spectrum is broad, noisy, aliased, or represented by only a small number of pulses. Autoregressive and maximum-entropy spectrum estimators can improve spectral resolution for short records, but they introduce model-order and stability choices and do not directly give the finite-sample Gaussian Doppler moment likelihood \cite{Palmer1991,Klostermeyer1989MaximumSignals}. Maximum-likelihood spectral estimators provide a stronger statistical model and can approach finite-sample performance bounds, but their practical use has been limited by the cost of repeatedly evaluating a likelihood over many range gates \cite{Levin1965PowerEstimation,Dias1993MaximumRatios}. This tradeoff is especially important for modern fast-scanning radars, where the dwell time is short and the number of coherent pulses per beam can be small.}

Maximum-likelihood type estimators are attractive in this setting because they use an explicit data model. They can account for the finite number of pulses, receiver noise, and the expected Doppler spectral shape instead of relying only on a few sample moments. In weather-radar practice, however, such estimators are rarely used as the main processor because a radar volume contains many range gates, azimuths, elevations, and updates in time. A useful maximum-likelihood Doppler estimator must therefore be more than accurate in simulation: the likelihood itself must be cheap to evaluate, must have analytic derivatives, and must be initialized from quantities that are already available in an operational processor.

The PSE method in \cite[Secs. III-IV]{10336781} addresses the accuracy problem by fitting a finite-sample semianalytical Doppler spectrum to the measured periodogram. It improves the moment estimates, but its direct implementation is slower than PP and DFT because every range gate requires repeated Whittle likelihood evaluations. This computational gap is the main obstacle for using PSE in an operational radar processor.

This letter focuses on that obstacle and develops an operational Whittle maximum-likelihood form of PSE for short weather-radar dwell times. The contribution is threefold. First, the Gaussian Doppler model is expressed through closed-form autocorrelation lags that directly include the finite record length. Second, the finite-sample PSE spectrum and its derivatives are synthesized by FFTs using precomputed normalized lag factors. Third, the estimator is extended to estimate total power jointly with mean Doppler frequency and spectral width. In an operational processor, PP and DFT estimates can still be used as seeds and as a practical way to decide where the fast PSE should be applied, but this gate-selection step is a deployment recommendation rather than the main algorithmic contribution evaluated here. \revone{The numerical validation covers several SNR values and both spectral-width and mean-Doppler sweeps.} \revtwo{The mean-Doppler sweep includes off-center normalized frequencies close to the Nyquist edge, and the accuracy results use $N=64$, representative of short coherent dwell processing in scanning weather radars.}

With this structure, each likelihood evaluation can be organized as
\begin{equation}
\theta \rightarrow \phi_m(\theta) \rightarrow F_k(\theta)
\rightarrow {\cal L}(\theta), \nabla {\cal L}(\theta),
\end{equation}
with no numerical integration and no finite differences. Here $\theta=[\mu_{fn},\sigma_{fn},P]^T$ contains normalized mean Doppler frequency, normalized spectral width, and total signal power; $\phi_m(\theta)$ is the Doppler autocorrelation at lag $m$; $F_k(\theta)$ is the finite-sample model spectrum at DFT bin $k$ including noise; and ${\cal L}(\theta)$ is the Whittle negative log-likelihood (NLL) \cite{Whittle1953} of the averaged periodogram.

The main body of the paper is organized as follows. Section II presents the signal model, the Whittle likelihood, and the FFT derivatives. Section III presents the Monte Carlo and timing results, Section IV discusses operational use, and Section V concludes the letter.

\section{Fast Finite-Sample PSE}
\subsection{Signal and Periodogram Model}
For the accuracy study, we generate the weather-radar echo as a complex Gaussian time series with the desired Doppler moments. For pulse indices $n$ and $r$, the signal covariance is
\begin{equation}
R_s[n-r] =
P \exp\{j2\pi \mu_{fn}(n-r)\}
\exp\{-2\pi^2\sigma_{fn}^2(n-r)^2\}.
\label{eq:covariance}
\end{equation}
The simulated time series is obtained by a Cholesky factorization of the covariance matrix and independent circular complex Gaussian samples. If $\mathbf{R}_s=\mathbf{L}\mathbf{L}^{H}$, then one realization is generated as
\begin{equation}
\mathbf{x}=\mathbf{L}\mathbf{u}+\sigma_n\mathbf{e},
\qquad
\mathbf{u},\mathbf{e}\sim{\cal CN}(\mathbf{0},\mathbf{I}),
\label{eq:chol_generation}
\end{equation}
where $\sigma_n^2$ is the receiver-noise power.

All Doppler quantities in this letter are normalized before estimation. If $T_s$ is the pulse repetition interval and $f_D$ is the Doppler frequency in hertz, then $f_{n}=f_D T_s$ is the Doppler frequency in cycles per pulse. Equivalently, for radial velocity $v$, $f_n=2vT_s/\lambda$. Thus $\mu_{fn}\in[-0.5,0.5)$ is the normalized mean Doppler frequency and $\sigma_{fn}$ is the normalized Doppler spectral width. This normalization is important computationally: for a fixed $N$ and Doppler grid, all range gates use the same bounded parameter domain and the same finite-sample lag dictionary. In this implementation, the model and the periodogram are evaluated on the $N$-point DFT grid, i.e., $K=N$, with no zero padding or oversampling.

Let $\hat{Z}_k$ be the measured averaged periodogram on this $N$-point normalized Doppler grid. The NLL used by PSE follows the periodogram likelihood formulation in \cite[Sec. IV]{10336781} and is written here as
\begin{equation}
{\cal L}(\theta)=
\sum_{k=0}^{N-1}
\left[
\log F_k(\theta)+\frac{\hat{Z}_k}{F_k(\theta)}
\right],
\label{eq:whittle}
\end{equation}
where $F_k(\theta)$ is the finite-sample model spectrum including receiver noise.

For a Gaussian Doppler spectrum, the normalized Doppler autocorrelation at lag $m$ is
\begin{equation}
\phi_m(\theta) =
P \exp(j2\pi m\mu_{fn})
\exp(-2\pi^2m^2\sigma_{fn}^2).
\label{eq:lagmodel}
\end{equation}
The finite record length $N$ applies a triangular lag weighting, as in the finite-sample PSD derivation of \cite[Sec. III, App. A]{10336781}. With
\begin{equation}
a_m = 2\left(1-\frac{m}{N}\right), \quad m=1,\ldots,N-1,
\end{equation}
the finite-sample model can be written as
\begin{equation}
F_k(\theta) = \sigma_n^2 + P\,G_k(\mu_{fn},\sigma_{fn}),
\label{eq:model}
\end{equation}
\begin{align}
G_k ={}& 1+\Re\left\{
\sum_{m=1}^{N-1}
a_m \exp(j2\pi m\mu_{fn})\right. \nonumber\\
&\left.{}\times
\exp(-2\pi^2m^2\sigma_{fn}^2)
\exp(-j2\pi km/N)
\right\}.
\label{eq:gk}
\end{align}
Equation \eqref{eq:gk} is the same finite-sample model used in the direct PSE synthesis \cite[Sec. III, App. A]{10336781}, but it is now expressed as a Fourier synthesis of lag-domain samples.

\subsection{Analytic Gradient}
The derivative of \eqref{eq:whittle} is
\begin{equation}
\frac{\partial {\cal L}}{\partial \theta_i}=
\sum_{k=0}^{N-1}
\left(
\frac{1}{F_k}
-\frac{\hat{Z}_k}{F_k^2}
\right)
\frac{\partial F_k}{\partial \theta_i}.
\label{eq:grad}
\end{equation}
After computing $F_k$, define
\begin{equation}
w_k=\frac{1}{F_k}-\frac{\hat{Z}_k}{F_k^2}.
\end{equation}
Then the gradient is a weighted inner product between $w_k$ and the model-spectrum derivatives. The required lag derivatives are closed form:
\begin{align}
\frac{\partial F_k}{\partial P} &= G_k,\\
\frac{\partial F_k}{\partial \mu_{fn}} &=
P\Re\left\{\mathcal{F}
\left[
a_m(j2\pi m)q_m
\right]\right\}_k,\\
\frac{\partial F_k}{\partial \sigma_{fn}} &=
P\Re\left\{\mathcal{F}
\left[
a_m(-4\pi^2m^2\sigma_{fn})q_m
\right]\right\}_k,
\end{align}
where
\begin{equation}
q_m=\exp(j2\pi m\mu_{fn})\exp(-2\pi^2m^2\sigma_{fn}^2)
\end{equation}
and $\mathcal{F}\{\cdot\}$ denotes the FFT over the Doppler bins. Thus, each likelihood and gradient evaluation only needs a few FFTs and vector operations.

The computational gain comes from separating fixed radar-processing quantities from the unknown Doppler moments. For a fixed coherent processing length $N$, the lag index $m$, the finite-sample weights $a_m$, the $N$-point FFT ordering, and the noise floor are independent of the range gate and can be stored once. The normalized variables make this practical because $\mu_{fn}$ always lies in the Nyquist interval and $\sigma_{fn}$ lies in a bounded operational interval. If this normalized search space is quantized, the factors $\exp(j2\pi m\mu_{fn})$, $\exp(-2\pi^2m^2\sigma_{fn}^2)$, and their derivative factors can also be stored for the allowed $\mu_{fn}$ and $\sigma_{fn}$ candidates. Then a likelihood call only selects or interpolates these lag-domain factors, multiplies them by the current power, and applies an $N$-point FFT to obtain $F_k$ and its derivatives. This is the main difference from the direct PSE implementation, where the finite-sample spectrum is rebuilt repeatedly by dense summation.

\subsection{Operational Version}
The fast PSE remains an off-grid estimator, but the computation is grid assisted. For a fixed $N$, the normalized search domain for $\mu_{fn}$ and $\sigma_{fn}$ is bounded; therefore $m$, $a_m$, FFT ordering, and optional phase/width dictionaries for candidate normalized values can be stored once. PP and DFT provide physically meaningful initial points. At each trial $\theta$, the cached phase and width factors form the finite-sample lag vector, $F_k(\theta)$ and its derivatives are synthesized by FFTs, and the NLL and analytic gradient are passed to the bounded optimizer. The total power is estimated jointly with $\mu_{fn}$ and $\sigma_{fn}$, or profiled inside the local search when this reduces the numerical dimension. This structure keeps the estimator off-grid while exploiting the shared normalized lag representation across range gates.
For reproducibility, the bounded local searches in the MATLAB implementation use \texttt{fmincon} from the Optimization Toolbox \cite{MathWorksOptimizationToolbox}. The direct three-parameter PSE is run with supplied objective gradients and bound constraints on $\mu_{fn}$, $\sigma_{fn}$, and $P$; the fast version uses the same bounded parameterization with the FFT-based NLL and analytic gradient, initialized from PP/DFT estimates.

Table \ref{tab:complexity} summarizes the computational change for the $K=N$ implementation used in this paper. The direct form evaluates the finite-sample spectrum by dense lag-to-frequency synthesis and normally uses numerical or finite-difference derivatives. The fast form uses the same model, but replaces dense synthesis by $N$-point FFTs and uses analytic gradients.

\begin{table}[t]
\caption{Per-Likelihood Evaluation Cost for One Range Gate}
\label{tab:complexity}
\centering
\begin{tabular}{lll}
\toprule
Method & Model synthesis & Derivatives \\
\midrule
Direct PSE & $O(N^2)$ & finite differences or dense sums \\
Fast PSE & $O(N\log N)$ & analytic FFT derivatives \\
PP/DFT & $O(N)$ or $O(N\log N)$ & not iterative \\
\bottomrule
\end{tabular}
\end{table}

\section{Numerical Study}
\revone{The simulation uses $N=64$ pulses and the same $K=64$ Doppler bins for all methods, true power $P=10^4$, and SNR values of $0$, $10$, $20$, and $30$ dB. Two parameter sweeps are considered. In the first, $\mu_{fn}=0$ is fixed and $\sigma_{fn}$ is swept from $0.01$ to $0.4$. In the second, $\sigma_{fn}=0.11$ is fixed and $\mu_{fn}$ is swept from $-0.45$ to $0.45$. At each grid point, independent realizations are generated from \eqref{eq:chol_generation}. We compare DFT, PP, direct PSE, and fast PSE. The retrieval error is summarized by root-mean-square error (RMSE) for the three estimated parameters, so that bias and variance contributions are visible in one metric. The finite-sample CRB from \cite[App. B]{10336781} is also shown as the locally unbiased variance reference.}

\revtwo{Scanning weather radars commonly use short coherent dwell lengths with roughly $N=16$-$64$ pulses per beam. The accuracy results use $N=64$, and the timing comparison reports operational values at $N=16$, $32$, and $64$. The $N=512$ point is included only as a complexity-scaling check for the FFT implementation.}

\begin{figure*}[!t]
\centering
\includegraphics[width=\textwidth]{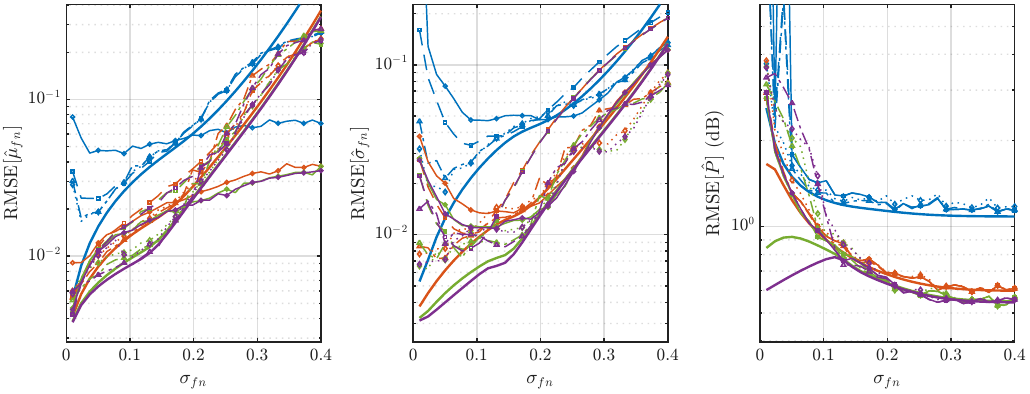}\\[-1mm]
\includegraphics[width=0.7\linewidth]{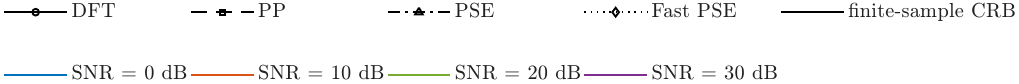}
\caption{\revone{Monte Carlo RMSE versus normalized spectral width for several SNR values, with $N=K=64$, $\mu_{fn}=0$, and $P=10^4$. The columns show normalized mean Doppler frequency, normalized spectral width, and total power in dB. The finite-sample CRB is overlaid as the locally unbiased variance reference.}}
\label{fig:revision_sigma_snr}
\end{figure*}

\begin{figure*}[!t]
\centering
\includegraphics[width=\textwidth]{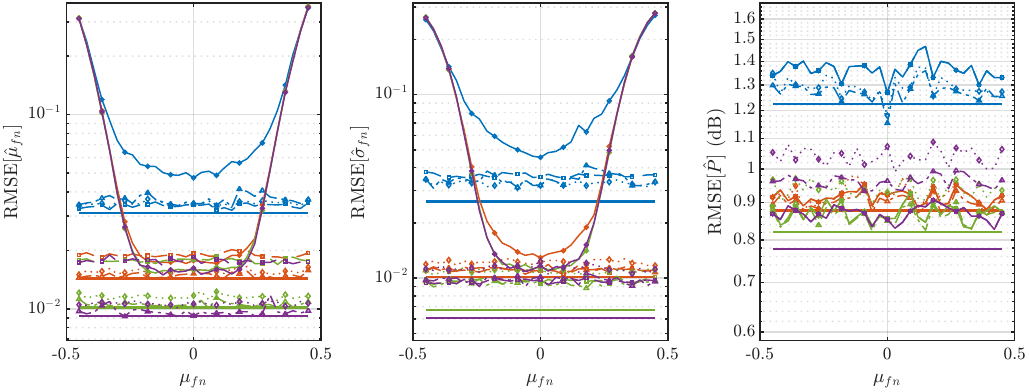}\\[-1mm]
\includegraphics[width=0.7\linewidth]{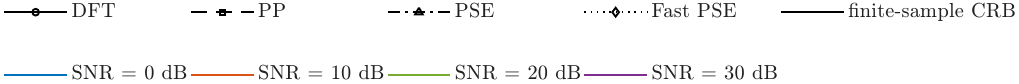}
\caption{\revtwo{Monte Carlo RMSE versus normalized mean Doppler frequency for several SNR values, with $N=K=64$, $\sigma_{fn}=0.11$, and $P=10^4$. The sweep includes off-center Doppler frequencies close to the Nyquist edge. The finite-sample CRB is overlaid as the locally unbiased variance reference.}}
\label{fig:revision_mu_snr}
\end{figure*}

\revone{Fig.~\ref{fig:revision_sigma_snr} shows that the fast PSE preserves the accuracy behavior of the direct PSE across the SNR range. For broad spectra, DFT and PP exhibit increasing spectral-width error, whereas the direct and fast PSE reduce this error because they fit the finite-sample Doppler PSD model, which inherently accounts for the limited resolution. Since the plotted error is RMSE, the CRB should be read as the unbiased variance reference and not as a biased-estimator RMSE bound.}

\revtwo{Fig.~\ref{fig:revision_mu_snr} evaluates off-center mean Doppler frequencies. Near the Nyquist edge, DFT and PP become more sensitive to wrapping and finite-sample distortion, while the fast PSE remains close to the direct PSE.}

\revone{We have also applied these techniques to real weather radar data. The measured-data example uses clutter-suppressed HH-polarized MESEWI X-band data acquired at TU Delft on 9 May 2023. For $N=64$, measured data from three range-azimuth are processed with DFT, PP, direct PSE, and fast PSE. The resulting moments are plugged into the same finite-sample semianalytical spectrum $F_k(\theta)$ and compared with the measured periodogram on the native dB power scale.}

\begin{figure*}[!t]
\centering
\includegraphics[width=0.98\textwidth]{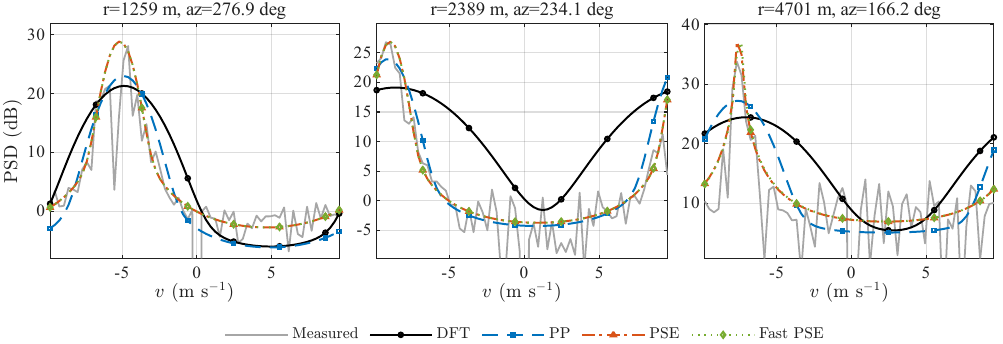}
\caption{\revone{Measured MESEWI spectra and finite-sample semianalytical reconstructions for the three range-azimuth cells with $N=64$. Range is in meters, and azimuth is in degrees. The same method styles as in Fig.~\ref{fig:revision_sigma_snr} are used. The direct and fast PSE curves overlap because they minimize the same NLL up to optimizer tolerance.}}
\label{fig:real_data_reconstruction}
\end{figure*}

\revone{Fig.~\ref{fig:real_data_reconstruction} checks measured-data consistency over multiple range-azimuth cells: the accelerated implementation reaches the direct-PSE likelihood solution, while PP and DFT remain low-cost references and initializers.}

\begin{figure}[!t]
\centering
\includegraphics[width=\columnwidth]{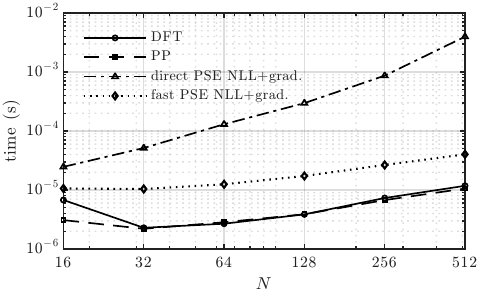}
\caption{\revtwo{Single-call MATLAB timing versus coherent processing length. The points $N=16$, $32$, and $64$ represent usual short-dwell scanning-weather-radar cases; $N=512$ is shown only as a complexity-scaling check. The fast PSE uses the FFT-based NLL and analytic gradient, while the direct PSE uses dense finite-sample synthesis.}}
\label{fig:timing_scaling}
\end{figure}

\revtwo{The timing comparison was made per single range-gate estimate. Fig.~\ref{fig:timing_scaling} shows the objective-level scaling: the NLL and gradient evaluation is $2.3$, $5.0$, and $10.4$ times faster at the usual short-dwell values $N=16$, $32$, and $64$, respectively. The $N=512$ value, where the objective-level speedup is about $95$ times, is included only to verify the expected large-$N$ change from dense $O(N^2)$ synthesis to FFT-based $O(N\log N)$ synthesis in Table \ref{tab:complexity}. Including initialization and optimizer overhead, the median end-to-end MATLAB times for $N=16$, $32$, and $64$ were $1.39$, $2.08$, and $3.49$ ms for the fast PSE, compared with $11.3$, $8.75$, and $9.58$ ms for the direct PSE. These full-estimator timings remain in the millisecond range but are not expected to scale monotonically because they include optimizer stopping behavior and fixed MATLAB overhead.}

\section{Discussion}
The proposed implementation changes how PSE is computed. The finite-sample correction from \cite[Sec. III, App. A]{10336781} remains in the model, the total received power is estimated jointly with mean and width, and the likelihood is still the Whittle likelihood used for averaged periodograms. The main practical difference is that the repeated likelihood evaluations are now simple FFT and vector operations.

For an operational weather radar, the processor can compute PP/DFT moments everywhere, flag difficult gates, and run fast PSE only where its bias reduction is needed.

\section{Conclusion}
This letter presented a fast three-parameter Whittle maximum-likelihood PSE for weather-radar Doppler moments. \revtwo{FFT synthesis and analytic gradients reduce each NLL and gradient evaluation from dense $O(N^2)$ summation to $O(N\log N)$ operations, giving measured objective-level MATLAB speedups of \textbf{2.3}, \textbf{5.0}, and \textbf{10.4} times for the usual operational dwell lengths $N=16$, $32$, and $64$, respectively. The corresponding fast-PSE median end-to-end times are \textbf{1.39}, \textbf{2.08}, and \textbf{3.49} ms per range gate, while the direct-PSE times are \textbf{11.3}, \textbf{8.75}, and \textbf{9.58} ms. The $N=512$ case gives an objective-level speedup of about \textbf{95} times and is used only as a complexity-scaling check.} Monte Carlo and measured-data examples show that the fast implementation preserves the direct-PSE behavior while making maximum-likelihood Doppler moment estimation more practical for operational radar processing.

{\scriptsize
\bibliographystyle{IEEEtran}
\bibliography{references}
}

\end{document}